\documentclass{eptcs}
\usepackage{amsmath, amssymb, wrapfig}
\usepackage{graphicx}
\usepackage{url}
\usepackage{listings}
\usepackage{longtable}
\usepackage{multirow}
\usepackage{array}
\usepackage{graphicx}
\usepackage{IEEEtrantools}

\title{Analyzing Consistency of Behavioral REST Web Service Interfaces}
 \author{Irum Rauf, Ali Hanzala Khan and Ivan Porres
\institute{\AA bo Akademi University, Dept. of Information Technologies,Turku, Finland}
\email{firstname.lastname@abo.fi}}
\def\titlerunning{Analyzing Consistency of Behavioral REST Web Service Interfaces}
\def\authorrunning{I. Rauf, A.H. Khan \& I. Porres}

\usepackage{color}

\newcommand{\DONE}[2][$\Delta$]{}

\usepackage{url}

\usepackage{graphicx}
\usepackage{listings}
\newenvironment{definition}[1][Definition]{\begin{trivlist}
\item[\hskip \labelsep {\bfseries#1}]}{\end{trivlist}}
 \newcommand{\code}[1]{\texttt{#1}}

\begin{document}
 \maketitle

\begin{abstract}
REST web services can offer complex operations that do more than just simply creating, retrieving, updating and deleting information from a database. We have proposed an approach to design the interfaces of behavioral REST web services by defining a resource and a behavioral model using UML. In this paper we discuss the consistency between the resource and behavioral models that represent service states using state invariants. The state invariants are defined as predicates over resources and describe what are the valid state configurations of a behavioral model. If a state invariant is unsatisfiable then there is no valid state configuration containing the state and there is no service that can implement the service interface. We also show how we can use reasoning tools to determine the consistency between these design models.
\end{abstract}

\section{Introduction}

The design phase of a software development lifecycle is crucial in the development of a reliable software, since the design models developed in this phase are carried forward to all the other phases. It is therefore important that these design models are constructed correctly. The design models are created from different viewpoints to capture different features of the system under development, since all the features are difficult to capture in a single model and can make a single model complex. Capturing the system under development in different models from different viewpoints gives a better and simpler understanding of the system, but raises the issue of models inconsistency. Models can become inconsistent, if they define the same system but have contradicting specifications in different models or have specifications that cannot be satisfied and hence its implementation can have undesirable results. This raises the need for consistency analysis of design models. The need for consistency analysis of models can rise even if the models themselves have no errors. Designers may specify certain requirements in different models that contradict each other and thus, cannot exist together leading to inconsistent diagrams. These mistakes can lead to implementations that do not provide correct functionality as expected from them.

As the software shifts from software as a product to software as a service, the need for consistency analysis of design models rises even further since the service users have no control over the software and completely rely on advertised service specifications. Also, since web services are offered from remote locations that consumers use via Internet using standard internet protocols, they can be expensive in terms of bandwidth and other costs. Thus, it is important to deliver services that are reliable and do not contain unintended design mistakes to avoid undesired results.

In this paper, we discuss a consistency checking approach for the design models of behavioral REST web service interfaces. REST~\cite{fthesis} is an architectural style to design scalable web services which play well with the existing infrastructure of web. They usually offer simple interfaces that can create, retrieve, update and delete information from database. However, it is possible to create REST web service interfaces that do more than simple CRUD operations. Such beyond CRUD REST interfaces offer different service states that need to be preserved as the consumer goes through trails of resources.

In \cite{sacporresnrauf}, we present a design approach that creates behavioral REST web service interfaces by construction. Designing and publishing a REST web service interface with stateful behavior may involve many resources and different service states that are dependent on these resources. This can result in inconsistencies leading to services implementations with unintended behavior. In this paper we present a consistency checking approach that analyzes the REST design models to detect inconsistent behavior and as such advises the developer to correct the detected design mistakes and create consistent behavioral REST web service interfaces. A behavioral interface is said to be consistent if it does not contain any contradicting specifications and there exists a service that can satisfy it.

The paper is organized as follows. Section 2 provides an overview of the REST design models and highlights the inconsistency problem. Section 3 defines the problem of determining the consistency of REST interface design models and explains our reasoning approach. Section 4 formally defines the structure of REST design models and Section 5 addresses description logic and OWL~2. In Section 6 we explain the translation of design models in OWL~2 ontology and Section 7 discusses the consistency analysis using an OWL~2 reasoner. Related work is presented in Section 8 and Section 9 concludes the paper.

\section{REST Design Models and their inconsistencies}

REST is a resource-centric architecture and a REST interface exposes resources that can be manipulated using standard HTTP methods. It offers features of \emph{connectivity}, \emph{addressability}, \emph{statelessness} and \emph{uniform interface}. Connectivity requires that there is no isolated resource and every resource is reachable. Addressability feature requires that every resource can be reached independently with a URI (path). The statelessness feature requires that no hidden session or state information is passed in the method calls and the uniform interface feature requires that the same set of methods (standard HTTP methods) is used to manipulate all resources. A behavioral REST interface should exhibit all these features of a REST interface and also provide information on how to use a service, e.g. sequence of method invocations and effects of service requests on the service.

\subsection{Modeling REST behavioral interfaces }
We represent a behavioral REST interface with resource and behavioral models using UML class and protocol state machine diagrams from UML~\cite{uml2009ss}, respectively, with some additional constraints to make them RESTful. For example, Figure~\ref{fig:hb_cm} and Figure~\ref{fig:hb_bmninv} shown in Section 4 depicts a resource model and a behavioral model for a hotel booking REST web service interface that takes payment from the customer and books a room in the hotel. The web service reserves a room for the customer and uses a third party payment service for confirmation. The service can be canceled when it is not processing payment and can be deleted only if it is canceled.The example is simple to understand and helps in demonstrating complex service states.

A resource is a piece of information that is exposed via a URI and that can be manipulated with standard HTTP methods. A resource can be either a collection resource or a normal resource. Collection resource does not have any attributes of its own and contains a list of other resources, whereas, a normal resource has its own attributes and represents a piece of information. In our resource model, we represent \emph{resource definitions} as classes, such that instance of these \emph{resource definitions} are termed as resources, analogous to the relationship between \emph{class} and its \emph{objects} in object oriented paradigm. A collection resource definition is represented by a class with no attributes and a normal resource definition has one or more attributes. Each association has a name and minimum and maximum cardinalities. These cardinalities define the minimum and maximum number of resources that can be part of the association. We also define a root resource definition in the resource model that represents the service. Root resource definition is connected to every other resource definition in the behavioral model. In Figure~\ref{fig:hb_cm}, \emph{Booking} is the root resource definition.

The behavioral specifications of a REST interface are represented as a behavioral model that represents different states of a service during its lifecycle, the methods that can be invoked on these resources and a sequence of these methods invocations. A state in a behavioral model represents the resource configuration of the service at a particular instance of time. A transition (from source state to target state) with a trigger method indicates the change of service state when an HTTP method with a side effect is invoked. The only allowed methods in our behavioral model are HTTP GET, PUT, POST and DELETE methods leading to a uniform interface. Of these methods, GET is idempotent and does not change state of the service whereas PUT, POST and DELETE have side-effects and can change state of the service.

\subsection{Linking Resource and Behavioral Models and Inconsistency Problems}

Each service state has a state invariant. We define invariants of states as predicates over resources defined in the resource model and that can have either true or a false value. For a state to be active, its state invariant should be true, otherwise it should be false.   When the client makes a service request, it is mapped to a transition in the behavioral model that has that method as a trigger. The transition is fired from a source state to a target state. If the state invariant of source state is inconsistent, a service can never exist in this state and it would be impossible for the implementation of the interface to decide which transition to take as a result of a service request. For example in Figure~\ref{fig:hb_bmninv},  the state invariant of state \emph{processingPayment} is $self.payment->size()=1~and~payment.waiting = True$. If we change its invariant to have $payment.waiting = False$, then it could conflict with the state invariant of $unpaidBooking$, i.e. both the states can be true at the same time. In this case, if PUT is invoked on $payment$ resource, the implementation would not know which transition to take. Such inconsistency problems can lead to service implementations with undesirable behavior.

\section{Consistency Analysis}
\label{sec:ClassAndSM}
In this section we define the problem of determining the consistency of our REST web service design models. Our view of model consistency is inspired by the work of Broy et al.~\cite{BCGR09a}. This work considers the semantics of a UML diagram as their denotation in terms of a so-called system model and defines a set of diagrams as consistent when the intersection of their semantic interpretation is nonempty.

In our work, we assume that there is a nonempty set ${\rm \Delta}^{\mathcal I}$ called the domain containing all the possible resources and resource configurations in our domain. We propose that a design model depicting a number of resource and behavioral diagrams is interpreted as a number of subsets of ${\rm \Delta}^{\mathcal I}$ representing each resource definition and each state in the model and as a number of conditions that need to be satisfied by these sets.

A resource definition is represented by a set $R$, such that $R \subseteq {\rm \Delta}^{\mathcal I}$. A resource r belongs to a resource definition R iff $r \in R$. We also represent each state S in a statechart as a subset of our domain $S \subseteq {\rm \Delta}^{\mathcal I}$. In this interpretation, the state set $S$ represents all the resources in the domain that have such state active, that is, resource r is in state S iff $r \in S$.

Since resource and behavioral models are represented with class and state machine diagrams respectively, other elements that can appear in a UML model such as generalization of classes, association of classes, state hierarchy and state invariants are interpreted as additional conditions over the sets representing resources and states. For example specialization is interpreted as a condition stating that the set representing a subresource is a subset of the set representing its superresource. These conditions are described in detail in the next section.

In this interpretation, the problem of design model consistency is then reduced to the problem of satisfiability of the conjunction of all the conditions derived from the model. If such conditions cannot be satisfied, then a design model will describe one or more resource definitions that cannot be instantiated into resources or resources that cannot ever enter a behavioral state in a statechart. This can be considered a design error, except in the rare occasion that a designer is purposely describing a system that cannot be realized.

\subsection{Reasoning Tool Chain}
\label{sec:approach}
In order to determine the satisfiability of the concepts represented
in our design model, we propose to represent the resource and behavioral models using a
Description Logic, and analyze the satisfiability of the concepts using
an automated reasoning tool. We have chosen OWL~2 DL to represent our
UML models since we consider it is well supported and adopted, and
there exist several OWL~2 reasoners for checking concept
satisfiability.

A
number of resource models, behavioral models and state invariants are
taken as an input. All the inputs are translated to the OWL~2 DL, a web
ontology language~\cite{owlrec}. The OWL~2 translation of design models
are passed to a reasoner. The reasoner provides report of
unsatisfiable and satisfiable concepts. Unsatisfiable concepts will
reveal resource definitions that cannot be instantiated or behavioral states that
cannot be entered.

We have also implemented tool that generates a) skeleton of REST web services from design models, b) OWL~2 DL from design models. The generation of REST web service skeleton is implemented using python in Django web framework~\cite{djangobook}. We translate resource and behavioral models into models.py, urls.py and views.py that are basic files of Django.  The skeleton for each resource contains information on its representation and its relative URI, the allowed methods, and the preconditions and postconditions for methods that have side-effects. The developer can inspect the generated code and fill in the desired logic for methods in the generated REST web service skeletons. The tool that generates OWL~2 DL from design models is discussed later in Section~\ref{sec:tool}.

In order to translate the design models into OWL~2 ontology, we need to first formally define the structure of our design models.

\section{Structure of Behavioral RESTful Interfaces}
\label{lbl:str}

In this section, we formally define the resource and behavioral models of behavioral REST interfaces and define the constraints that make them RESTful.

\subsection{Structure of Resource Model}
\begin{figure*}[t]
\centering
\includegraphics[scale=0.5]{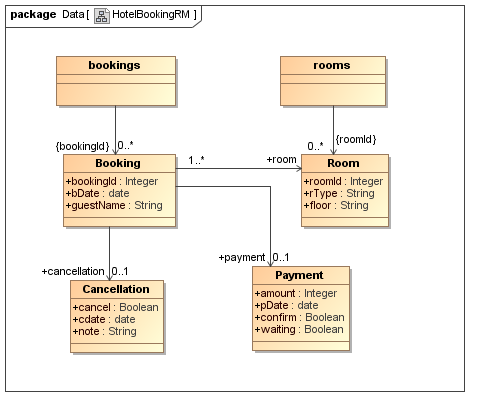}
\caption{Resource Model for HB RESTful Web Service with Invariants}
  \label{fig:hb_cm}
\end{figure*}
The structure of a REST interface is represented with a class diagram from UML~\cite{uml2009ss} with some additional constraints. Figure~\ref{fig:hb_cm} shows the resource model of Hotel Booking REST web service interface. It has six resources with two collection resources: bookings and rooms, and four normal resources: \emph{Booking}, \emph{Room, Cancellation }and \emph{Payment}. There is no isolated resource definition and every resource definition is connected via association. Each association has a name and multiplicity constraint.
\\
\begin{definition}{\bfseries 1}
A resource model RM is defined as a tuple: $RM =  \langle R\_def , root, Att, A, l, issubresource,$ $ min, max \rangle $, where $R_\_def$ is a set of resource definitions, $root$ is the root resource definition and $Att$ is a set of attributes. Also:
\label{def_cm}
\end{definition}

\begin{itemize}
\item $A$ is a set of associations where each association connects two resource definitions.  We define association as a relation between two resource definitions, i.e. $a: R\_def \times R\_def$.
\item $l(a)$ gives name of the association. $l$ is defined as an injective function from set of associations A to L, the set of labels, i.e. $l: A \rightarrow L$. The injective function implies that every association has a unique name.
\item $min(a)$ and $max(a)$ give the minimum and maximum cardinality on association $a$. They are both defined as a function from set of associations to natural numbers, i.e.,$min: A \rightarrow N$ and $max: A \rightarrow N$, such that $min(a) \le max(a)$.
\item $issubresource(r_1, r)$ evaluates to true if $r_1$ is a subresource of $r$. Resource model can have resource hierarchy in which subresources of a resource inherit the properties and attributes of its parent resource.

\end{itemize}

$R\_def$ represents \emph{resource definition} that defines a resource in resource model. The instances of these resource definitions are resources. The relation between \emph{resource definition} and its \emph{resources} is same as to the relationship between \emph{class} and its \emph{objects} in object oriented paradigm where class represents all the entities that share same set of properties and its \emph{object} represents its instance.

The set of resource definitions $R\_def$ in RM is the union of a collection resource $R_c$ and a normal resource $R_n$ definitions, i.e. $ R\_def = R_c\_def \cup R_n\_def $ and $R_c\_def \cap R_n\_def = \emptyset$
\begin{itemize}
\item $R_c\_def$ is a set of collection resource definitions. A collection resource does not contain any attribute of its own, i.e. $R_c\_def = \{r_c \in R_c\_def: \forall r_c \in R_C \land \forall att \in Att: att \notin r_c\} $, where $r_c$ is an instance of collection resource $R_c$.
\item $R_n\_def$ is a set of normal resource definitions. We call a resource normal (not collection) if a resource has at least one attribute, i.e., $R_n\_def = \{r_n \in R_n\_def:  \exists att \in Att: att \in r_n \} $
\end{itemize}

We take a \emph{root} as a resource definition that represents the service. All other resource definitions are linked to \emph{root} and are navigated through it. In Figure~\ref{fig:hb_cm}, we take \emph{Booking} resource definition as \emph{Root}. This root resource definition is accessed with its specific booking Id, i.e., the starting navigation path for all the resource definitions in resource model is $/\{bookingId\}/$.

In order to exhibit features of connectivity and addressability, we constrain our resource model with the following design decisions.
\subsubsection{Connectivity }

All resource definitions should be connected via associations and every resource definition is reachable from the root resource definition such that there is no isolated resource definition or sub-graph. For example, the resource model in Figure~\ref{fig:hb_cm} is connected because there is a path from root resource definition $Booking$ to every other resource definition.
%

\subsubsection{Addressability }

The addressability feature of REST interface requires that every resource should have a URI address. We can retrieve the relative navigation path to a resource definition from resource model by concatenating the association names of associations that make a path to the resource definition. For example, the payment resource definition in Figure~\ref{fig:hb_cm} can be reached via $/\{bookingId\}/payment$. We ensure the addressability feature by constraining each association to have a unique association name and direction. This is implied by the injective $l$ function defined above. These association names and directions give the addressable path to the resource definitions and navigation directions.

%

\subsection{Structure of Behavioral Model}
 \begin{figure*}[t]
\centering
\includegraphics[scale=0.47]{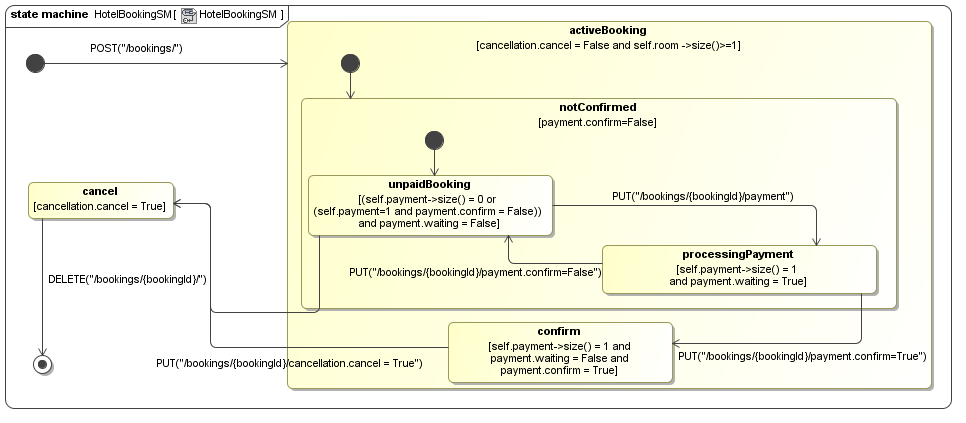}
\caption{Behavioral Model of HB REST Web Service Interface}
  \label{fig:hb_bmninv}
\end{figure*}
The behavior of a REST interface is represented as a protocol state machine from UML~\cite{uml2009ss} with some additional constraints. Figure~\ref{fig:hb_bmninv} shows the behavioral model of Hotel Booking REST web service interface. A \emph{Booking} resource is created when POST is called on \emph{bookings} collection resource. A \emph{Booking} is active and unpaid when it is created. When user of the service invokes a PUT on \emph{payment} resource, the web service goes to \emph{processingPayment} state. If the payment is confirmed to be true, it goes to \emph{confirm} state, otherwise it goes back to the \emph{unpaidBooking} state. The booking cannot be cancelled if it is processing the payment. When DELETE is invoked on \emph{Booking} resource from state \emph{cancel}, it is deleted from the system.
\begin{definition}{\bfseries 2}
A behavioral model of REST web service is given as a tuple:

      $BM =\langle S, \iota, F, T, \sigma, g, pc, inv, issubstate, trigger, region \rangle$

 where  S is a set of states, $\iota$ is the initial state such that $\iota \in S$, F is the set of final states such that F $\subseteq$ S, T is a set of transitions and $\sigma$ gives the resource configuration of the web service in a particular state such that $\sigma \in \sum$ where $\sum$ is the set of all possible resource configurations. Also,
 \end{definition}
  \begin{itemize}
    \item $g(\sigma,t)$ evaluates the guard and $pc(\sigma,t)$ evaluates the postcondition of transition t in state $\sigma$ . They evaluate to true in case they are not mentioned.
    \item $inv(s, \sigma)$ evaluates the invariant of state s of a service in state $\sigma$.  The invariant of state s for a RESTful web service is retrieved by invoking GET method on all resources that are part of resource configuration in $\sigma$ in state s.
    \item $issubstate(s, s_1)$ evaluates to true if $s$ is a substate of $s_1$.
    \item $trigger(t)$ gives the trigger method for transition t and is defined as a function, $trigger: T        \rightarrow Trigger$, where $Trigger =\{PUT, POST, DELETE\}$
    \item $region(s_1, s_2)$ is a predicate that returns true if state $s_1$ and $s_2$ belong to the same region.
  \end{itemize}

We make the following main design decisions in the construction of our behavioral interface for REST web service to address features of REST interface.
\subsubsection{Uniform Interface}

We only allow HTTP methods GET, PUT, POST and DELETE in our behavioral model. The GET method is idempotent and has no side-effects. GET method is used to retrieve the state of the resources that constitute state invariants. The only allowed methods that can trigger a state change in our state machine are PUT, POST and DELETE.

%

\subsubsection{Statelessness }

When a method is invoked, it makes a transition from one service state to another. The statelessness feature requires that no hidden state or session information should be passed as part of method call such that each HTTP request is treated as an independent request. This features leads to scalability of web services. For this feature we require that all the information passed with the method call is either part of URL or is in request parameters.

Also, in order to define statelessness of REST interface, we define states of a service as predicates over resources. A state is active when the resource configuration defined in its state invariant is true otherwise false. We, thus, define service states of a REST web service without violating the statelessness feature of REST interface.

\section{Description Logic and OWL 2}
\label{sec:DLandOWL2}
The Description Logic used in our approach is classified as $SROIQ$~\cite{horrocks06}. Description Logic is made up of concepts, denoted here by $C,D$, and roles, denoted here by $R,Q$. A concept or role can be named, also called atomic, or it can be composed from other concepts and roles.

An interpretation $\mathcal I$ consists of a non-empty set
$\rm\Delta^{\mathcal I}$ and an interpretation function which assigns a set
$C^{\mathcal I}\subseteq {\rm \Delta}^{\mathcal I}$ to every named concept $C$ and a binary
relation $R^{\mathcal I} \subseteq {\rm \Delta}^{\mathcal I} \times {\rm
\Delta}^{\mathcal I}$ to every named role $R$.

The constructors of Description Logic are as follows:
\begin{IEEEeqnarray*}{rClrCl}
\top^{\mathcal I} & = & {\rm \Delta}^{\mathcal I} & (\forall R.C)^{\mathcal I} & = & \{ x \mid \forall y. (x,y) \in R^{\mathcal I}
\to y \in C^{\mathcal I} \} \\
\perp^{\mathcal I} & = & \emptyset & (\exists R.C)^{\mathcal I} & = & \{ x \mid \exists y . (x,y) \in R^{\mathcal
I} \land y \in C^{\mathcal I} \} \\
(\  neg C)^{\mathcal I} & = & {\rm \Delta}^{\mathcal I} \backslash C^{\mathcal
I} & (\geq n\, R)^{\mathcal I} & = & \{ x \mid \#\{ y \mid (x,y)\in R^{\mathcal I} \}
\geq n \} \\
(R^{-})^{\mathcal I} & = & \{ (y,x) \mid (x,y) \in R^{\mathcal I} \} & \hspace{0.2cm}(\leq n\, R)^{\mathcal I} & = & \{ x \mid \#\{ y \mid (x,y)\in R^{\mathcal I} \}
\leq n \} \\
(C \sqcap D)^{\mathcal I} & = & C^{\mathcal I} \cap D^{\mathcal I} &
(C \sqcup D)^{\mathcal I} & = & C^{\mathcal I} \cup D^{\mathcal I} \\
\end{IEEEeqnarray*}

where $\# X$ is the cardinality of $X$. Axioms in DL can be either inclusions $C \sqsubseteq D$,$C \sqsubseteq D$ or equalities $C \equiv D$,  $R \equiv Q $.


An interpretation satisfies an inclusion $C \sqsubseteq D$ if $C^{\mathcal I}
\subseteq D^{\mathcal I}$ and an inclusion $R \sqsubseteq Q$ if $R^{\mathcal
I} \subseteq Q^{\mathcal I}$. An interpretation satisfies an equality $C
\equiv D$ if $C^{\mathcal I}
= D^{\mathcal I}$ and an equality $R \equiv Q$ if $R^{\mathcal
I} = Q^{\mathcal I}$. $\mathcal I$ satisfies a set of axioms if it
satisfies each axiom individually -- $\mathcal I$ is then said to be a model
of the set of axioms. Given a set of axioms $\mathcal K$, a named concept $C$ is said to be
satisfiable if there exists at least one model $\mathcal I$ of $\mathcal K$ in which
$C^{\mathcal I} \neq \emptyset$. A set of axioms is said to be satisfiable if
all of the named concepts that appear in the set are satisfiable. If a set of
axioms $\mathcal K$ is satisfiable, we say that an axiom $\phi$ is satisfiable
(with respect to $\mathcal K$) if ${\mathcal K} \cup \{ \phi \}$ is
satisfiable. Similarly, we say that $\phi$ is unsatisfiable (w.r.t. $\mathcal
K$) if ${\mathcal K} \cup \{\phi\}$ is unsatisfiable.

The decidability of  $SROIQ$ is demonstrated by Horrocks et al.~\cite{horrocks06} and there exist several reasoners that can process answer satisfiability problems automatically~\cite{pellet,hermit,Tsarkov:2006:FDL:2136107.2136140}.

\subsection{OWL 2 Functional Syntax}
For practical reasons, we use the OWL 2 functional syntax (OWL2fs)~\cite{owlrec} as the language used as an input for the reasoners and in the text of this article. The interpretation of the main OWL 2 expressions used in this article is presented in the following table. A complete description of the semantics OWL 2, including support for data types can be found in~\cite{OWL2ds}.

\begin{center}
\begin{tabular}{ll}
\texttt{SubClassOf(C1 C2)} & ${C_1} \sqsubseteq {C_2}$  \\
\texttt{EquivalentClasses(C1 C2)} &  ${C_1} \equiv {C_2}$  \\
\texttt{DisjointClasses(C1 C2) } & ${C_1} \sqcap {C_2} = \emptyset$  \\
\texttt{ObjectPropertyDomain(P C)} &  $\forall R^{-1}.C $ \\
\texttt{ObjectPropertyRange(P C) } & $\forall R.C $\\
\texttt{ObjectMinCardinality( n P) } & $\geq n\, R$ \\
\texttt{ObjectMaxCardinality( n P) }& $\leq n\, R$ \\
\texttt{ObjectExactCardinality(n P)}  & $(\geq n\, R) \sqcap (\leq n\, R)$ \\
\end{tabular}
\end{center}

In the next section, we discuss and translate the structure of a resource and behavioral model with state invariants over the sets representing resource definitions and states into OWL~2 DL.

\section{From Resource and Behavioral Diagrams to OWL~2 DL}
\label{sec:modelstoOWL2}

 In order to check the satisfiability of resource definitions in a resource model, we need to first translate all resource definitions and their associations into OWL~2 ontology, and then validate the OWL~2 ontology using an OWL~2 reasoner. In this section we only present the translation of those concepts of resource model which are required for the validation of the behavioral model such as: resource definitions, association, multiplicity and attributes.
\subsection{Resource Model in OWL~2}
Each resource in a conceptual model is shown as a class in an ontology and an association as an object property.  A class in OWL~2 is a set of individuals and \emph{ObjectProperty} connect pair of individuals\cite{owlrec}. According to the definition of a resource model given in Definition 1, we need to map these concepts in OWL~2 DL: resource definitions and their specializations, attributes, associations and association multiplicities.

\subsubsection{Resource Specification and Hierarchy}

A resource definition in a resource model represents a collection of resources which share same features, constraints and definition. For each resource definition ${R\_def}$ in \emph{RM}, we define an OWL~2 axiom: \code{Declaration(Class(R\_def))}

$issubresource(r_1, r)$ is true if $r_1$ is a subresource of resource $r$. We explicitly define the hierarchy of resources in OWL~2 between resources. The specialization of resources represented as classes is reduced to the set inclusion. We represent the fact that a resource definition R1 is a specialization of resource definition R2 with the condition $R1 \subseteq R2$. In this case we say that R2 is a super resource of R1, analogous to superclass in UML class diagram. If two resource definitions R1 and R2 have a common super resource, or R2 is the super resource of R1 we say that they are in a specialization relation. The specialization relation $R_1 \subseteq R_2$ is translated in OWL~2 as:
\begin{verbatim}
SubClassOf( R1 R2 )
\end{verbatim}

Each resource definition at the same hierarchical level in resource model represents a different piece of information. We assume that a resource cannot belong to two resource definitions, except when these two resource definitions are in a specialization relation. In our semantic interpretation of a resource diagram, it is equally important to denote the facts that two resource definitions are not in a specialization relation. We represent the fact that two resources R1 and R2 are not in a specialization relation with the condition $R1 \cap R2 = \emptyset$. With this condition, a resource cannot belong to these two resource definitions simultaneously. Due to the open-world assumption used in Description Logic, we need to explicitly state this fact in OWL~2, i.e. for resource definitions R1...Rn at same hierarchical level we define disjointness in OWL~2 as:
\begin{verbatim}
DisjointClasses( R1..Rn )
\end{verbatim}

\subsubsection{Attributes}
In our resource model, a collection resource does not have any attribute and a normal resource should have at least one attribute. So we define attribute $att$ of a normal resource $r$ of type $D$ as $DataProperty(att)$ with domain as $r$ and range as $D$. We do not need to give any attribute definition for collection resources because OWL~2 has open world assumption and that which is not mentioned is not considered. So by simply not mentioning collection resources with any attributes is sufficient. However, for every normal resource, each of its attributes is defined in OWL~2. Attributes usually have a multiplicity restriction to one value. Hence, the attribute definition \emph{att} in OWL~2 is given as:
\begin{verbatim}
Declaration(DataProperty( att ))
SubClassOf(C DataExactCardinality(1 att ))
DataPropertyDomain( att r )
DataPropertyRange( att D )
\end{verbatim}

\subsubsection{Association}

An association is a relation between two resource definitions, $r_1$ and $r_2$ and name of association is its label \emph{l}. For each association $a$ in A, we define \code{ObjectProperty} axiom with label \emph{l} and $r_1$ as its domain and $r_2$ as its range,i.e., for $a(r_1, r_2)$ with $l(a)$, we give OWL~2 definition as:
    \begin{itemize}
        \item $l$ maps to OWL~2 axiom \code{Declaration(ObjectProperty(l))}
        \item $r_1$ maps to OWL~2 axiom \code{ObjectPropertyDomain(l $r_1$)}
        \item $r_2$ maps to OWL~2 axiom \code{ObjectPropertyRange(l $r_2$)}
    \end{itemize}

$min(a)$ and $max(a)$ give minimum and maximum cardinality of association $a$. It defines the number of allowed resources that can be part of the association. We represent a directed binary association A from resource definition R1 to R2 as a relation $A:R_1xR_2$.  The multiplicity of the association defines
additional conditions over this relation $\#\{y|(x,y) \in A \} \geq
min$, $\#\{y|(x,y) \in A \} \leq max$. If an association $a$ has minimum cardinality $min$ and maximum cardinality $max$ for resource r, we define it in OWL~2 as:
\begin{verbatim}
SubClassOf( r ObjectMinCardinality( min a ) )
SubClassOf( r ObjectMaxCardinality( max a ) )
\end{verbatim}

\subsection{Behavioral Model in OWL~2}
A behavioral model provides the behavioral interface of a web service and defines the sequence of method invocations, the conditions under which they can be invoked and their expected results. To check the satisfiability of state invariants in a behavioral model, we need to translate the states and their invariants into OWL~2. The translation of the state and the state invariant includes the reference of resources and their attributes so we translate a behavioral model in the same ontology that contains the OWL~2 translation of a resource model.

We need to cover following concepts of our behavioral model in OWL~2: state, state hierarchy, state disjointness and state invariant.


\subsubsection{State and State Hierarchy}
 A state in behavioral model is a concept that defines the state of the service. We represent a state of behavioral model as a set of resources that have that state
active. A set representing a state will be a subset of the set
represented with the root resource associated to the behavioral model,
since all the resources that can have the state active belong to the
root resource $r$. That is, if the state $S$ belong to a state machine
associated to root resource $r$, then $S \subseteq r$. We represent this in OWL 2
as follows:
\begin{verbatim}
Declaration(Class(S))
SubClassOf( S r )
\end{verbatim}
State hierarchy is also represented using set inclusion. Whenever a
substate is active, its containing state is also active. This implies
that if a resource belongs to a set representing the substate will
also belong to the set representing the super state, $sub \subseteq S$.
We define each substate relationship explicitly in OWL~2. For each state $sub$ that is a substate of composite state $s$, i.e. $issubstate(sub,s)$, we define OWL~2 axiom as:
\begin{verbatim}
SubClassOf( sub s )
\end{verbatim}

The states at same hierarchical level represent different resource configurations such that only one state can be active at same time. The composite states with non-orthogonal regions also follow the same principle, i.e. only one state can be active at same hierarchical level.  This means that a resource cannot be at
the same time in the two sets representing two exclusive states, i.e.
if $S_1$ and $S_2$ represents substates of an active and not orthogonal
composite state then $S_1 \cap S_2 = \emptyset$. When representing a
state machine in OWL 2, the non-orthogonal exclusive states are declared as disjoint, so that they may not able to share any resource.
\begin{verbatim}
DisjointClasses( S1..Sn )
\end{verbatim}
In a composite state with orthogonal regions, two or more states can be active at the same time if they belong to two or more different regions of composite state, i.e, if $R_1$ and $R_2$ are the regions of an active and orthogonal composite state $S$  then $R_1 \cup R_2 = S$. We should note that if $region(s_1) \neq region(s_2)$  then they are not exclusive and $S_1 \cap S_2 \neq \emptyset$. Due to the open-word assumption of DL we do´not need to define this non-exclusiveness since classes may represent same set of instances unless they are explicitly declared as disjoint.

\subsection{State invariant into OWL~2 DL}

The invariant condition characterizes the state, i.e., if
the invariant condition holds the state is active, otherwise if the
invariant condition does not hold the state is not active.

In our approach we represent an invariant as a set of resources that makes that
invariant evaluate to true. Since the invariant holds iff the
associated state is active, the set representing a state will be the
same as the set representing an invariant. This is represented in  OWL~2 as an equivalent class relation between the state and its invariant:
\begin{verbatim}
EquivalentClasses(S Invariant)
\end{verbatim}
Due to the equivalent relationship between state and its invariant,
all resources that fulfill the condition of its state invariant will
also be in that specific state.

\subsubsection{State Constraints}
Our behavioral model is represented by a UML protocol state machine with additional constraints. The UML allows us to define additional constraints to a state,
and names these constraints also state invariants. However, the
semantics of a state constraint is more relaxed since it ``specifies
conditions that are always true when this state is the current state''
(\cite{UMLSE241}, p.562). In this sense, the state constraints define
necessary conditions for a state to be active, but not sufficient.
This means that, the actual state invariant may remain implicit. However, we consider a state invariant as a predicate
characterizing a state. That is, a state will be active if and only if
its state invariant holds.

The UML Superstructure specification requires that whenever a state is
active its state invariant evaluates to true (\cite{UMLSE241},p.562).
A consequence of this is that state invariants should be satisfiable.
That is, every state invariant in a state machine must hold in at
least one resource configuration. Otherwise there cannot be resources that
have such state active. Since invariants should be satisfiable, the
set of resources $S$ representing a state should not be empty $S \neq
\emptyset$.

\subsection{State Constraints in $\mu$ OCL}
\label{sec:OCLtoOWL2}
A state invariant is a runtime constraint on the state (\cite{UMLSE241},p.514). It is used to express a number of constraints, such as the restriction on the values of resource attributes or the restriction on the existence of resources by using the multiplicity constraint of the associations. These constraints are combined by using boolean operators. We have used a subset of OCL to define state invariants in state machine diagrams. In this section, we will discuss and translate the different types of OCL constructs supported in our approach.

\subsubsection{Attribute Constraints}
The value of the attribute is accessed in OCL by using a keyword $self$ or by using a class (resource) reference (\cite{OMG_OCL2},p.15), the value constraint of the attribute $Att$ is written in OCL as \texttt{self.Att=Value}, meaning $\{ x | ( x , Value ) \in Att \}$, where $Value$ represents the attribute value. The OCL attribute value constraint $self.Att=Value$ is mapped to OWL~2 axiom $DataHasValue$ as follows:
\begin{verbatim}
DataHasValue(Att "Value"^^datatype )
\end{verbatim}
where $Att$ is the name of the attribute, $Value$ is the value of the attribute, and $datatype$ is the datatype of the attribute $Value$.

\subsubsection{Multiplicity Constraints}
 The multiplicity of an association is accessed by using $size()$ operation in OCL (\cite{OMG_OCL2},p.144). The multiplicity constraint on the association $A$ in OCL is written as \texttt{self.A-$>$size()=Value}, where $Value$ is a positive integer and represents the number of allowable resources of the range resource definition of the association $A$. We can use a number of value restriction infix operators with $size()$ operation such as $=$, $>=$, $<=$, $<$ and $>$. The multiplicity constraint on an association $A$ is defined as $\{ x | \#\{ y | ( x , y ) \in A \} OP~Value \}$, where $OP$ is the infix operator and $Value$ is a positive integer. The translation of $size()$ operation in OWL~2 is based on the infix operator used with the $size()$ operation, such as:
\begin{itemize}
\item "$size()>=$" or "$size()>$" is mapped to OWL~2 axiom: $ObjectMinCardinality$
\item "$size()<=$" or "$size()<$" is mapped to OWL~2 axiom: $ObjectMaxCardinality$
\item "$size()=$" is mapped to OWL~2 axiom: $ObjectExactCardinality$
\end{itemize}
For example, the OCL constraint $self.A->size()=Value$, in which $A$ is the name of an association and $Value$ is a positive integer, is written in OWL~2 as:
\begin{verbatim}
ObjectExactCardinality(Value A)
\end{verbatim}

\subsubsection{Boolean Operators}
 The constraints in a state invariant are written in form of a boolean expression, and joined by using the boolean operators, such as "$and$" and "$or$" (\cite{OMG_OCL2},p.144).
\begin{itemize}
\item The binary "$and$" operator evaluates to true when both boolean expressions $Ex_1~and~Ex_2$ are true. In our translation this is represented by the intersection of the sets that represent both expressions $Ex_1 \cap  Ex_2$. This is represented OWL 2 as \texttt{ObjectIntersectionOf(Ex1 Ex2)}.
\item The binary "$or$" operator evaluates to true when at least one of the boolean expression $Ex_1~or~Ex_2$ is true.  In our translation this is represented by the union of the sets that represent both expressions $Ex_1 \cup  Ex_2$.  This is represented OWL 2 as \texttt{ObjectUniounOf(Ex1 Ex2)}
\end{itemize}

\section{ Consistency Analysis using an OWL 2 Reasoning Tool}
\label{sec:tool}

We have defined earlier the satisfiability of our design models in Sect.~\ref{sec:ClassAndSM} and~\ref{sec:DLandOWL2}. The consistency analysis of resource and behavioral models is reduced to the satisfiability of the conjunction of all the conditions derived from the model. In order to determine the satisfiability of the conditions represented in design models, we first translate the resource and behavioral models into an OWL~2 ontology, then use an OWL~2 reasoner to analyze the satisfiability of translated concepts.

To translate the resource and behavioral models into OWL~2 ontology, we have implemented the translations of resource and behavioral diagrams in OWL~2, discussed in Sect.~\ref{sec:modelstoOWL2}, in form of a translation tool. We have used Python programming language for the implementation of the prototype of the translation tool. The implemented translation tool allows us to automatically transform a resource and behavioral model into OWL~2 DL. The translator takes these models and $\mu$OCL state invariant as an input in the form of XMI. The XMI is generated by using a modeling tool, Magicdraw. The XMI generated by the modeling tool contains the source code of both resource and behavioral model in form of XML. While modeling in a modeling tool we have used $\mu$OCL to express the state invariants in a behavioral model. The state invariant written in $\mu$OCL is also part of a XMI generated by the modeling tool. Moreover, the output of an implemented translation tool is an ontology file, which contain the transformed resource model, behavioral model and state invariants in from of OWL~2 functional syntax.

After translating the design models and state invariants into OWL~2 ontology by using the implemented translation tool, we will validate the output ontology by using an OWL~2 reasoner. The OWL~2 reasoner analyzes different facts presented as axioms in the ontology and infers logical consequences from them. When we give our ontology to the reasoner, it generates satisfiability report indicating which concepts are satisfiable and which not. If the ontology has one or more unsatisfiable concepts, this means that the instance of any unsatisfiable concept will make the whole ontology inconsistent, consequently, an instance of the resource definition describing an unsatisfiable concept in a resource diagram will not exist, or resources will not enter in a state describing an unsatisfiable condition, and vice versa. However, the ontology generated by the translation tool is in OWL~2 functional syntax, therefore, the satisfiability of the translated concepts can be checked by using any OWL~2 reasoner which supports OWL~2 functional syntax.

\section{Related Work}

Consistency analysis and checking of design models has been studied by number of researchers in the past but in the area of web services it has not been researched very extensively, especially in the area of REST web services, we were unable to find any consistency checking approaches. However, in the area of consistency checking for web services following works are noteworthy.

Yin et al. use type theory to verify consistency of web services behavior in \cite{yinyin}. The paper addresses web services choreography. It analyzes structure of service behavior and uses extended MTT, which is a constructive type theory, to formally describe service behavior. The procedures of deductions are then given that verify the suitability between services along with discussion on type rules for subtype, duality and consistency of web services behavior.

In  \cite{robusttesting} and ~\cite{tsaiweichen}, Tsai et al. present a specification based robust testing framework for web services. The approach first uses an event driven modeling and specification language to specify web services and then uses a completeness and consistency (C $\&$ C) approach to analyze them. Based on these positive and negative test cases are generated for robustness testing. The approach assumes that web services are specified in OWL-S. The approach aims towards testing of web services but C$\&$C analysis is performed on OWL-S specification that was point of interest for us. The approach identifies missing conditions and events, whereas, our approach checks the structure of web services and validates implementation of service requests. In \cite{caomiao}, Xiaoxia verify the service oriented requirements using model checking. The service-oriented computer independent model is used to structure the requirements and then automated model checking is done to do completeness and consistency checking of requirements. It provides formal definition of completeness checking as a check that all the required service are included in model and does not give any specific consistency checking constraint except the requirement relation applied on an example.

In \cite{nentwich}, Nentwich et al. present a static consistency checking approach for distributed specifications. It describes xlinkit framework that provides a distribution-transparent language for expressing constraints between web service specifications. It provides semantics for constraints that allow the generation of hyperlinks between inconsistent elements. The implementation of xlinkit is done on light-weight web service using XML.

In \cite{wsarch}, Heckel et al. present a model-based consistency management approach for web service architectures. They advocate use of UML class and activity diagrams for modeling web services. The consistency problems are then identified in UML based development process. For each consistency problem a partial formalization into a suitable semantic domain is done and a consistency check is defined. The consistency problems identified include syntactic correctness and deadlock freedom. Based on these, those activity diagrams are identified that are relevant to consistency checks. These are partially translated to CSP which are then assembled in a single file and handed over directly to model checker. The paper outlines a good consistency management approach for web services architecture that needs concrete development of different steps defined in the paper.

\section{Conclusion}

 A REST interface can do more than simply creating, retrieving, updating and deleting data from a database. Designing behavioral interface for such web services that provide different states of the service and offer REST interface features is an interesting design challenge since it can involve many resources and resource configurations that define different states of the service. In this paper, we address how to analyze the consistency of design models that create behavioral REST interfaces. We check the consistency of resource and behavioral diagrams with state invariants using OWL~2 reasoners. The structure of both the diagrams is formally defined and translated to OWL~2 ontology. The ontology is given to an ontology reasoner that checks the ontology for any unsatisfiable concepts. The unsatisfiable concepts indicate the design errors that can cause undesirable behavior in the implementation of the service. The approach is automated as we provide prototype tools that generate web service skeletons and OWL~2 ontology from the design models. Also, thanks to existing OWL~2 reasoners the generation of satisfiability report for our OWL~2 ontology is also automated that is analyzed to check the consistency of design models.

 In our future work, we plan to take into account guards on transitions in the consistency analysis of our design models.

\bibliography{RESTOntVerif}
\bibliographystyle{eptcs}

\end{document}